\documentclass[aps,prd,reprint,superscriptaddress,nofootinbib]{revtex4-2}

\newcommand{\bmt}{\begin{pmatrix}}
\newcommand{\emt}{\end{pmatrix}}
\newcommand{\ba}{\begin{array}{c}}
\newcommand{\ea}{\end{array}}
\newcommand{\be}{\begin{equation}}
\newcommand{\ee}{\end{equation}}
\newcommand{\bea}{\begin{eqnarray}}
\newcommand{\eea}{\end{eqnarray}}

\newcommand{\bi}{\begin{itemize}}
\newcommand{\ei}{\end{itemize}}

\newcommand{\baz}{\begin{array}{cc}}
\newcommand{\besub}{\begin{subequations}}
\newcommand{\eesub}{\end{subequations}}

\newcommand{\mathsym}[1]{{}}

\newcommand{\D}{\displaystyle}

\newcommand{\bt}{\begin{tabular}}
\newcommand{\et}{\end{tabular}}

\newcommand{\benu}{\begin{enumerate}}
\newcommand{\eenu}{\end{enumerate}}

\def\a{\alpha}

\def\ve{\varepsilon}

\def\D{\Delta}

\def\q2 {q^2}

\def\bt{\begin{table}}
\def\et{\end{table}}

\usepackage{graphicx}
\usepackage{hyperref}
\hypersetup{
    pdfnewwindow=true,
    colorlinks=true,
    linkcolor=blue,
    citecolor=blue,
    filecolor=blue,
    urlcolor=blue
}
\usepackage{xcolor}
\usepackage{amsmath}

\begin{document}

\preprint{IPPP/25/56}

\title{Boomerang mechanism explaining the excess radio background}

\author{P. S. Bhupal Dev}
\email{bdev@wustl.edu}
\affiliation{Department of Physics and McDonnell Center for the Space Sciences, 
Washington University, St. Louis, Missouri 63130, USA}
\author{Pasquale Di Bari}
\email{P.Di-Bari@soton.ac.uk}
\affiliation{School of Physics and Astronomy, University of Southampton, Southampton, SO17 1BJ, U.K.}
\author{Ivan Martinez-Soler}
\email{ivan.j.martinez-soler@durham.ac.uk}
\affiliation{Institute for Particle Physics Phenomenology, Department of Physics, Durham University, Durham DH1 3LE, UK}
\author{Rishav Roshan}
\email{r.roshan@soton.ac.uk}
\affiliation{School of Physics and Astronomy, University of Southampton, Southampton, SO17 1BJ, UK}


\begin{abstract}
We propose a {\em boomerang mechanism} for the explanation of the excess radio background detected by ARCADE 2. 
In an early stage of the Universe, at a temperature $T$ in the range $\sim 0.1\,{\rm keV}$--$1\,{\rm MeV}$, 
a fraction of relic neutrinos is resonantly converted 
into dark neutrinos by mixing induced by a pre-existing lepton asymmetry. Dark neutrinos 
decay much later into a dark-standard photon state and a dark fermion, with a lifetime longer 
than the age of the Universe, as required by a solution to the excess radio background. 
This scenario circumvents the upper bound on the neutrino magnetic moment but still 
implies a testable lower bound. 
\end{abstract}

\maketitle

\section{Introduction}
\label{sec:I}

\begin{figure}
\begin{center}
  \includegraphics[width=0.49\textwidth]{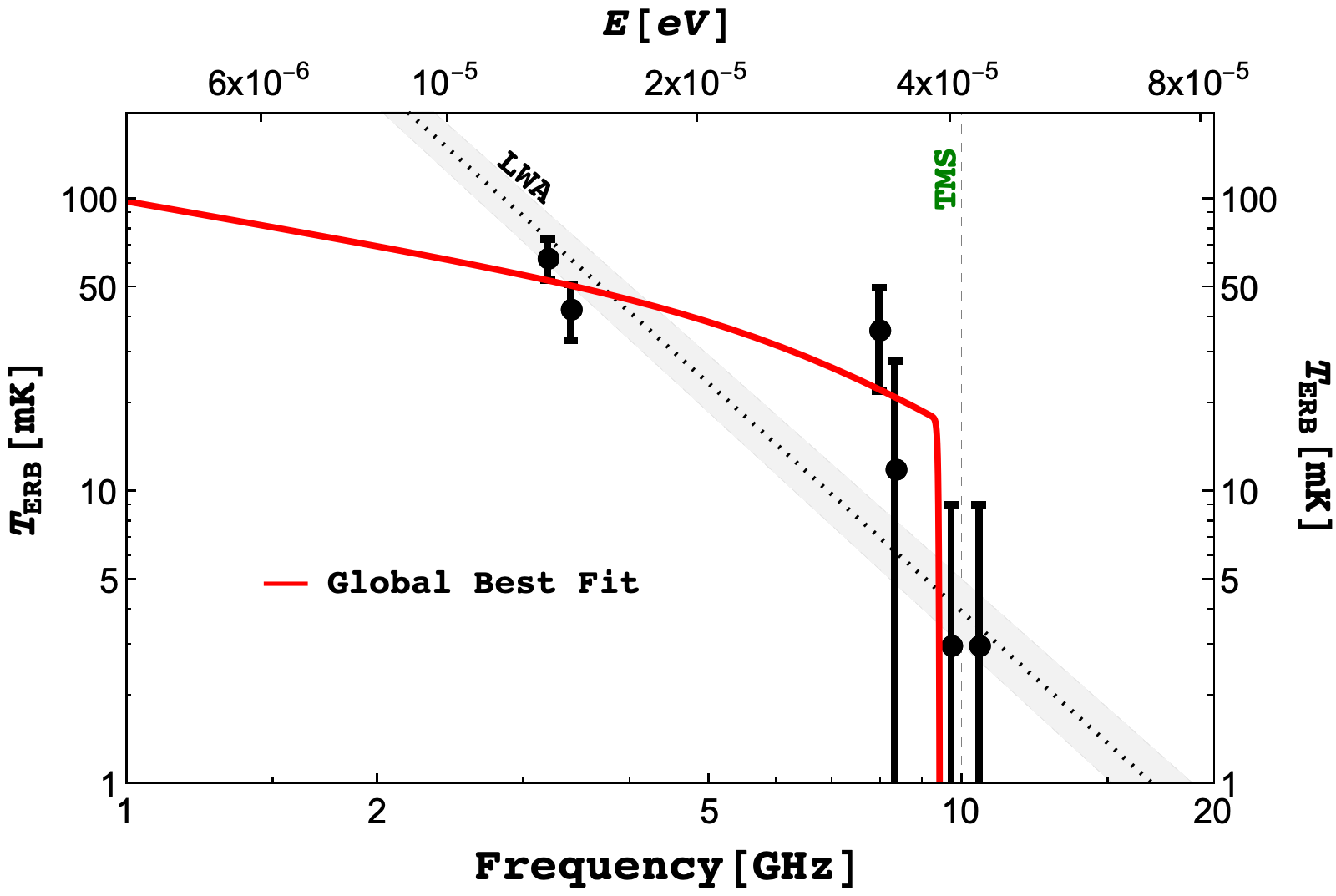}
\end{center}
    \caption{Best fit curve for $T_{\rm ERB}$ obtained with Eq.~(\ref{Tgammanth}). The thick red curve corresponds 
    to the best global fit obtained for $\D m_1 = 4.0 \times 10^{-5}\,{\rm eV}$  and $ \tau_1= 1.46 \times 10^{21}\,{\rm s}$.
    The ARCADE 2 data points are taken from Ref.~\cite{Fixsen:2009xn}. We also show the power-law fit 
    $\beta = -2.58 \pm 0.05$ (dotted line with grey shade), obtained using the Long Wavelength
Array (LWA) data at lower frequencies~\cite{Dowell:2018mdb}. The vertical dashed line shows the TMS low-frequency threshold.}
    \label{fig:bestfit}
\end{figure}

\begin{figure}[t]
    \centering
    \includegraphics[width=0.49\textwidth]{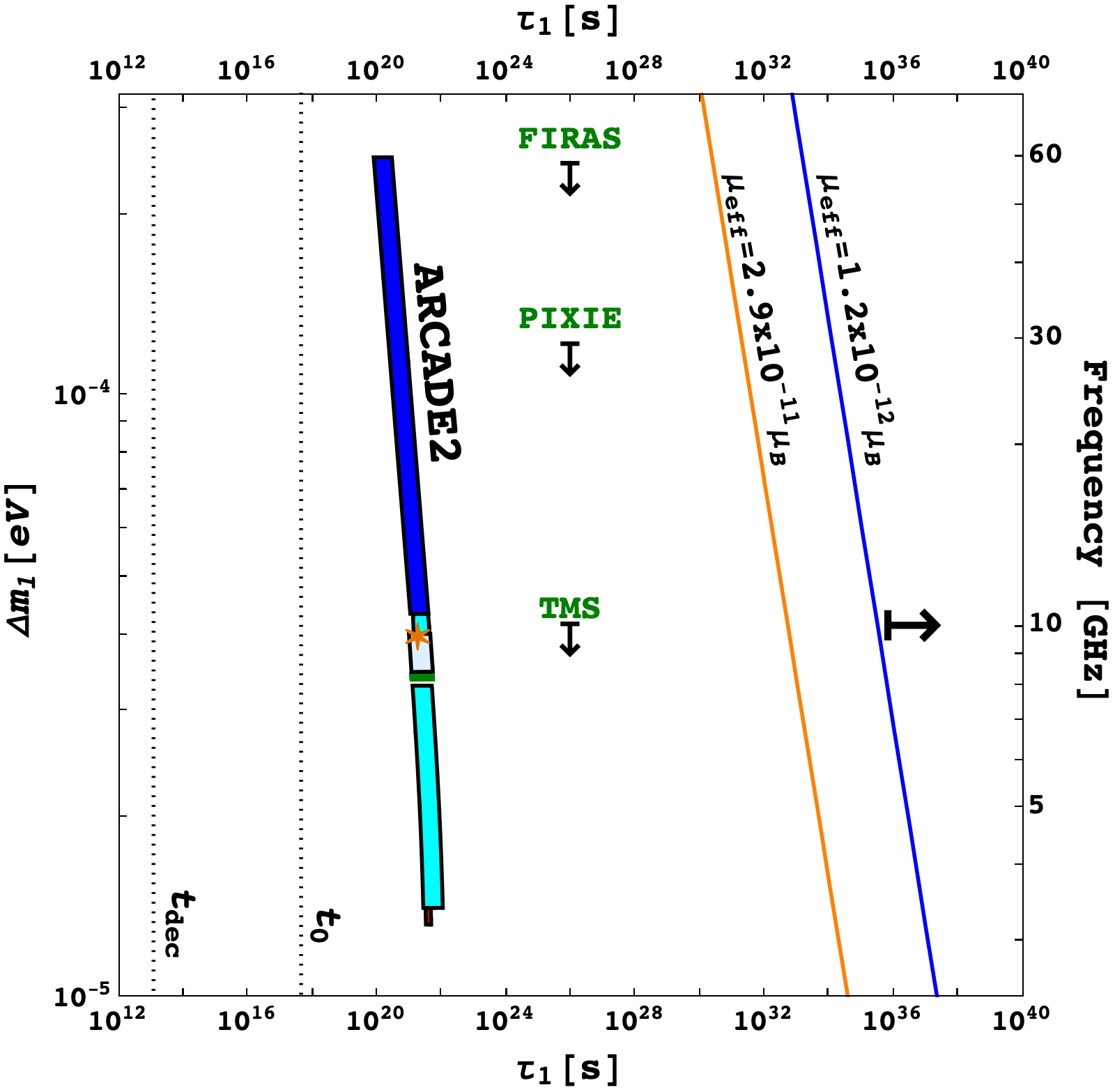}
    \caption{Allowed region, shaded with different shades corresponding to different photon energy bands 
     and with the best-fit point denoted by \textcolor{orange}{$\star$}, explaining the ARCADE 2 excess radio background  in the plane of $\D m_1$ vs. $\tau_1$~\cite{Dev:2023wel}. Lower bounds on the lifetime derived from the upper bound on the effective magnetic transition dipole moment [cf. Eq.~(\ref{Gammanu})] are shown by the blue and orange lines. 
    We also indicate the matter-radiation decoupling time $t_{\rm dec}$ and the current age of the Universe $t_0$. The lowest frequency thresholds for FIRAS, PIXIE and TMS are also indicated.}
    \label{fig:magnetic_moment}
\end{figure}

The Absolute Radiometer for Cosmology, Astrophysics and Diffuse Emission  (ARCADE 2)~\cite{Fixsen:2009xn}  
has detected an excess radio background (ERB) in the 3--10 GHz frequency range  with respect to the Cosmic Microwave Background (CMB) thermal spectrum. The excess is statistically significant (more than $5\sigma$) and  cannot be explained by known population of sources since they give a contribution to the effective temperature that is 3--10 times smaller than the measured one~\cite{Singal:2017jlh}.  Moreover, different observations  place a strong upper limit on the  anisotropy of the ERB that is, therefore, extremely smooth~\cite{Holder:2012nm}. This represents a strong constraint for an astrophysical origin and, therefore, the ERB is currently regarded 
as a mystery~\cite{Singal:2017jlh}.  The Tenerife Microwave Spectrometer (TMS) will  soon take data in the 10--20 GHz frequency range  
\cite{2020SPIE11453E..0TR,Alonso-Arias:2021quq} and might, therefore, help to shed light on this mystery. It was noticed that radiative relic neutrino decay 
can potentially explain the ARCADE 2 excess~\cite{Chianese:2018luo}. Recently, we have shown that  such a solution indeed fits very well the six
ARCADE 2 data points between 3--10 GHz giving rise to an excess~\cite{Dev:2023wel}. It predicts an effective (radiometric) temperature for the 
non-thermal photons produced by the decays of one of the relic neutrino species, the lightest for definiteness, given by
\be\label{Tgammanth}
T_{\gamma_{\rm nth}}(E,0) \simeq  {6\,\zeta(3)\over 11 \, \sqrt{\Omega_{{\rm M}0}}}\,
{T_{0}^3 \over E^{1 / 2}\, \Delta m_1^{3 / 2} } \, {t_0 \over \tau_1} 
\,\left(1 +{a_{\rm D}^3 \over a_{\rm eq}^3} \right)^{-{1 \over 2}} ,
\ee
where $t_0=4.35 \times 10^{17}\,{\rm s}$ is the age of the Universe, $T_0=(2.725 \pm 0.001){\rm K}$ is the photon temperature at the present time
measured by the Far Infrared Absolute Spectropho-
tometer (FIRAS) instrument~\cite{Fixsen:2009ug}, 
$\D m_1 \equiv m_1 - m_0 \ll m_1$ is the mass difference between the lightest active neutrino
and the new sterile state, $\tau_1$ is the neutrino lifetime, $E \leq \D m_1$ is the energy of the photon at the present time.\footnote{{\color{black}A  
discussion on the role of the absolute neutrino mass scale can be found in~\cite{Dev:2023wel}.}} 
For the definition and values of the other cosmological parameters in Eq.~\eqref{Tgammanth}, see Ref.~\cite{Dev:2023wel}.
This expression can be used to fit the six data points measured by ARCADE 2 in the 3--10 GHz frequency range. 
The best fit is obtained for $\D m_1 \simeq 4.0 \times 10^{-5}\,{\rm eV}$ and 
$\tau_1 \simeq 1.46 \times 10^{21}\,{\rm s}$. As
shown in Fig.~\ref{fig:bestfit}, it provides an excellent fit to the ERB temperature spectrum, improving a simple power-law fit,
with $\chi^2/{{\rm d.o.f.}} \simeq 1$ with 4 degrees of freedom (d.o.f.)~\cite{Dev:2023wel}. 
The result of the fit, expressed in terms of the quantity $\D m_1^{3/2}\tau_1$, gives at $99\%$ confidence level (C.L.):
\be\label{ARCADE}
(\D m_1^{3/2}\tau_1)^{\rm ARCADE} = 3.8^{+7.2}_{-1.5}  \times 10^{14}\,{\rm eV}^{3/2}\,{\rm s} \,  .
\ee
This region is shown in Fig.~\ref{fig:magnetic_moment} (shaded regions with different shades for different frequency bands) in the plane $\D m_1$ vs. $\tau_1$. 

In a general way, we can write the radiative decay rate of the neutrino mass eigenstate 
$\nu_1$ with mass $m_1$ into another neutrino mass eigenstate $\nu_0$ with mass $m_0$ 
in terms of the {\em effective magnetic transition dipole moment} $\mu_{{\rm eff}}$, as~\cite{Xing:2011zza}
\bea\label{Gammanu}
\Gamma_{\nu_1  \rightarrow \nu_0 +\gamma} & = & {(m^2_1 - m^2_0)^3 \over 8\pi \, m_1^3}\, \mu_{{\rm eff}}^2 \simeq {\Delta m^3_{1} \over \pi} \, \mu^{2}_{{\rm eff}}  \\ \nonumber
& \simeq & 42.5\,{\rm s}^{-1}\, \left( {m_1 - m_0 \over {\rm eV}}\right)^3 \,  \left({\mu_{{\rm eff}}\over \mu_{\rm B}}\right)^2 \, .
\eea
In the second expression we used $m_1 + m_0 \simeq 2\,m_1$, 
since we are interested in the case of quasi-degenerate neutrinos. In the numerical
expression we normalised $\mu_{{\rm eff}}$ to the Bohr magneton 
$\mu_{\rm B} = e\hbar/(2m_e) \simeq 296.3\,{\rm GeV}^{-1}$.
The effective magnetic transition dipole moment is defined as
\be
\mu_{{\rm eff}} \equiv \sqrt{|\mu_{1\rightarrow 0}|^2 + |\epsilon_{1\rightarrow 0}|^2} \,  ,
\ee 
where $\mu_{1\rightarrow 0}$ is the transition magnetic dipole and $\epsilon_{1\rightarrow 0}$ 
is the transition electric dipole moment \cite{Giunti:2014ixa}. 
The most stringent upper bound on $\mu_{{\rm eff}}$ comes from plasmon decays in globular cluster stars~\cite{Raffelt:1990pj, Capozzi:2020cbu}:
\be\label{upperbmu}
\mu_{{\rm eff}}  \lesssim 1.2\times 10^{-12}\,\mu_{\rm B}.
\ee
In this case, from Eq.~(\ref{Gammanu}), one easily finds the following lower bound on the lifetime of $\nu_1$:
\be
\tau_{\nu_1\rightarrow \nu_0 +\gamma} \gtrsim 2.5 \times 10^{21}\,{\rm s}\,
\left({{\rm eV}\over \Delta m_{1}} \right)^3 \,  .
\ee
One can see that for $\Delta m_{1} \lesssim 10^{-4}\,{\rm eV}$, a necessary condition to address
the ERB, one obtains 
$\tau_{\nu_1\rightarrow \nu_0 +\gamma} \gtrsim 10^{33}\,{\rm s}$, an incredibly long lifetime
yielding completely negligible contribution to the ERB. This conclusion remains valid
even considering the less stringent laboratory upper bound from the GEMMA experiment measuring electron recoils 
induced by neutrino-electron elastic scattering~\cite{Beda:2010hk,Beda:2012zz},
\be
\mu_{{\rm eff}}  \lesssim 2.9 \times 10^{-11}\,\mu_{\rm B} \,   .
\ee 
The situation is depicted in Fig.~\ref{fig:magnetic_moment}. 
Therefore, irrespectively of whether the final neutrino is an active or a sterile neutrino species, neutrino
radiative decay  by itself cannot give any sizable contribution to the ERB. 

The \emph{boomerang mechanism} that we propose here provides a way to circumvent this bound, preserving at the same time 
 the success of Eq.~(\ref{Tgammanth}) in reproducing the ARCADE 2 data. While in radiative neutrino decays ordinary neutrinos
 directly decay into photons, in the boomerang mechanism neutrinos are first converted into dark (sterile) neutrinos and then these 
 decay into dark photons and standard photons, with the latter constituting the ERB.\footnote{The 
 name `boomerang' mechanism is justified by the fact that the visible sector first throws sterile neutrinos into the dark sector via mixing, 
 and then these, decaying, throw back photons into the visible sector. The net result is that neutrinos are converted into photons.
 Trading off the effective magnetic moment of ordinary neutrinos with that of sterile neutrino to circumvent (\ref{upperbmu})
 is an ingredient, the only one, in common with the model proposed in Ref.~\cite{AristizabalSierra:2018emu} to explain the EDGES anomaly.}
 We first discuss in Section~\ref{sec:II} how active-dark neutrino mixing, in the presence of a sufficiently large lepton asymmetry, can 
 convert a fraction of relic neutrinos into (quasi-degenerate) dark neutrinos. 
 In Section~\ref{sec:III} we discuss how the dark neutrinos decay into a dark fermion species
 and into dark and standard photon state. In Section~\ref{sec:IV} we combine together all constraints and determine an allowed region
 in the plane of active-dark neutrino mixing angle versus mass squared difference that maps into a corresponding region
 in the plane of the neutrino effective magnetic moment and resonance temperature. In Section~\ref{sec:V} we compare our allowed region with current and future neutrino oscillation constraints. Finally,  we draw some final remarks and conclude in Section~\ref{sec:VI}. 

\section{Active-to-dark neutrino conversions in the early Universe}
\label{sec:II}

We assume the existence of a light sterile dark neutrino field almost coinciding with mass eigenstate $\nu_0$ with mass $m_0$
and quasi-degenerate with the lightest neutrino with positive $\Delta m^2 \equiv m^2_1 -m^2_0$.\footnote{We adopt a different
sign convention for $\Delta m^2$ than in~\cite{Enqvist:1990ad,DiBari:2001jk}.}  
We also assume,
for definiteness, that it just mixes with the muon neutrino but all results are valid in general. The much higher values of $m^2_{2,3}-m^2_0$ make in a way that the relevant mixing 
parameters are just $\Delta m^2$ and a very small muon-dark neutrino mixing angle $\theta_{0}$.  With these assumptions and definitions
the mixing in the early Universe is described by the effective Hamiltonian $\D H$ that in the interaction basis can be written as
\bea\label{effH}
& & \Delta H_{\mu-{\rm dark}}  =  
  \\ \nonumber
  & & {\Delta m^2\over 4\,p}  \, \left(
\begin{array}{cc}
\cos 2\theta_{0}- v(y,T_\nu,L) & -\sin 2\theta_{0} \\
-\sin 2\theta_{0} & -(\cos 2\theta_{0} - v(y,T_\nu,L)) \, 
\end{array}
\right) \,  ,
\eea
where we introduced the dimensionless effective potential 
\be
v(y,T,L) = v_1(y,T_{\nu}) + v_2(y,T_{\nu},L) \,  .
\ee
 In this expression we denoted by $T_{\nu}$ the neutrino temperature, 
$y\equiv p/T_{\nu}$ and by $L$ the effective (muonic) total asymmetry defined as
\be
L \equiv 2L_{\nu_\mu} + L_{\nu_e} +L_{\nu_\tau} -{1\over 2} B_n \,  ,
\ee
where the neutrino asymmetries $L_{\nu_\a} \equiv (N_{\nu_\a} - N_{\bar{\nu}_\a})/N^{\rm i}_{\gamma}$
and similarly for the neutron asymmetry $B_n$, with $N^{\rm i}_{\gamma}$ the photon abundance at
some initial time $t_{\rm i}$ such that $m_{\mu} \gg T_{\rm i} \gg m_e$, with $T_{\rm i} \equiv T(t_{\rm i})$. 
The term $v_1(y,T_{\nu})$ is the finite temperature contribution~\cite{Notzold:1987ik} and is given by
\be
v_1(y,T_{\nu}) = {{\rm eV}^2\over \Delta m^2}\left(T_{\nu}\over {\color{black}T_{\mu}}\right)^6 \,y^2 \,  ,
\ee
where $T_{\mu}\simeq 23.4~{\rm MeV}$. For $L=0$ and $\D m^2 >0$, there would be a resonance, both for neutrinos and antineutrinos,
at $T_{\nu}^{\rm res}(y,0) \simeq T_\mu\,(y^2 \, \D m^2/{\rm eV^2})^{1/6}$. At this resonance relic neutrinos are not efficiently converted
into dark neutrinos, though this could be used to trigger the generation of a large lepton asymmetry~\cite{Foot:1995qk,DiBari:2001jk}, 
as we discuss in subsection B. 

\subsection{Initial pre-existing effective asymmetry}
\label{sec:IIA}

If we assume that there already exists an initial pre-existing effective muon asymmetry $L_{\rm i}$,
 for example generated by the decays of weekly coupled seesaw neutrinos as in leptogenesis~\cite{Fukugita:1986hr}, 
one has also to consider the term $v_2(y,T_{\nu},L)$ given by~\cite{Enqvist:1990ad}
\be
v_2(y,T_{\nu},L) \simeq \mp \,v_0\,{{\rm eV}^2\over \Delta m^2}\,L \,\left({T_{\nu}\over {\rm MeV}}\right)^4\,y \,  ,
\ee
where the $-$ ($+$) sign holds for neutrinos (antineutrinos) and
$v_0 = (4\sqrt{2}\zeta(3)/\pi^2) 10^{12}\,G_{\rm F}{\rm MeV}^2 \simeq 8$.
If we assume that $|L_{\rm i}| \gg {L_{\star}(y)} \simeq 0.4 \times 10^{-6} (y \, \D m^2/{\rm eV^2})^{1/3}$, then
the resonance condition is  satisfied for $v_2(y,T_{\nu},L_{\rm i}) = 1$. This implies that 
for positive $L_{\rm i}$ and positive $\Delta m^2$ there is a resonance only for antineutrinos
at a resonant temperature 
\be\label{Tnures}
T_\nu^{\rm res}(y,L_{\rm i}\gg L_\star) = \left({1 \over v_0 \, L_{\rm i} \, y} {\Delta m^2 \over {\rm eV}^2}\right)^{1\over 4} {\rm MeV} < T_\nu^{\rm res}(y,0)   \,  .
\ee
Notice that this resonance occurs at different times for different values of $y=y_{\rm res}$. The asymmetry grows with time from the initial value
and $y_{\rm res}$ spans all neutrino distribution starting from a small value $y_{\rm res}\ll 1$ to large values $y_{\rm res} \gg 1$, 
as discussed in detail in Ref.~\cite{DiBari:2001jk}. Since this process occurs, in general, during electron-positron annihilations,
the neutrino temperature gets smaller than the photon temperature and, making use of entropy conservation,  one has
\begin{equation}
T_{\nu}=T\,\left[{g_S(m_e/T)\over g_S(0)}\right]^{1\over 3} \,  ,
\end{equation}
where $g_S$ is the number of entropy density ultrarelativistic degrees of freedom.
For $T_\nu^{\rm res}(y_{\rm res}\ll 1, L_{\rm i}) \lesssim 1\,{\rm MeV}$, neutrino collisions can be neglected and at the resonance one 
has antineutrino conversions into dark neutrinos  starting from small $y_{\rm res}\ll 1$. 
If the resonance is crossed adiabatically, then all lightest antineutrinos are converted into dark neutrinos. Notice that if $L_{\rm i}< 0 $, then simply 
lightest neutrinos are converted into dark neutrinos instead of antineutrinos. For definiteness, we will usually refer to the case $L_{\rm i}>0$ in the following.
More generally, to account also for non-adiabatic conversions, the fraction of converted neutrinos can be calculated using the Landau-Zener approximation:
\be
f_{\nu_\mu\rightarrow \nu_{\rm dark}} \simeq 1 - e^{-{\pi\over 2}\,\gamma_{\rm res}}  \,  .
\ee
In this expression $\gamma_{\rm res}$ is the adiabaticity parameter at the resonance, given by
\be\label{adpary}
\gamma_{\rm res} = {|\Delta m^2|\, \sin^2\,2\theta_0 \over 2y\, T_{\nu}^{\rm res}\,H_{\rm res}} \,  ,
\ee
where $H_{\rm res} \simeq 0.2{\rm s}^{-1}\, \sqrt{g_\rho(T_{\rm res})} (T_{\rm res}/{\rm MeV})^2$ 
is the expansion rate at the resonance and $g_{\rho}$ is the number of energy density ultrarelativistic degrees of freedom.
The Landau-Zener approximation has been shown to reproduce quite well  the numerical results obtained solving density matrix equation~\cite{DiBari:2001jk}.
For simplicity, we can make use of a monochromatic approximation equivalent to say
that all neutrinos are converted instantaneously at $T_{\nu}^{\rm res}$ corresponding at $y_{\rm res}=3.15$. 
Using the prescription in Ref.~\cite{DiBari:2001jk}, in this case one has to use $\langle L \rangle \simeq L_{\rm f}/2 \simeq 0.2$
in the evaluation of the adiabaticity parameter, obtaining:
\be\label{gammares}
\gamma_{\rm res} \simeq 1.4 \, \times 10^9 \, \sqrt{10.75 \over g_\rho(T_\nu^{\rm res})}\, \left({\Delta m^2 \over {\rm eV}^2} \right)^{1\over 4} \, \sin^2 2\theta_0 \,  .
\ee
A full momentum description would be able to describe the dependence of the adiabaticity parameter on momentum  
shown in Eq.~(\ref{adpary}). For values of $y_{\rm res}$ higher (lower) than $3.15$, the adiabaticity parameter would be smaller than
in the monochromatic approximation and so one can expect less (more) efficient conversion. However, bulk of neutrinos get converted
for $y_{\rm res} \simeq 3.15$ and one can expect that this value describes well the average effect. 
Therefore, an account of momentum dependence would produce just a slight correction, likely at the percent level, of the lower bound on the 
value of $\gamma_{\rm res}$ that in turn implies a lower bound on the active-dark neutrino mixing angle, as we derive in Section~\ref{sec:IV}.\footnote{Within the
current phenomenological status, such a small correction is very reliably negligible. If future upper bound on the neutrino effective magnetic moment will
be improved  by at least two orders of magnitude, then corrections of this kind might become worth to be included in the derivation of the allowed region.
At this stage a full momentum description is not really needed.}

\subsection{Dynamical generation of the effective asymmetry}
\label{sec:IIB}

The effective lepton asymmetry that is necessary to create a resonance at low temperatures
converting active neutrinos into dark neutrinos can also be generated dynamically by the 
same active-dark neutrino mixing~\cite{Foot:1995qk,DiBari:2001jk}. 
This second case is quite attractive, since in this way one does not have to assume additional external mechanisms. 
If $L_{\rm i}=0$, the initial resonance at $T_\nu^{\rm res}(0)\gtrsim 1\,{\rm MeV}$ occurs in the collisional regime and triggers an 
initial exponential growth of the asymmetry up to a threshold value $L_{\rm t} \simeq 0.4 \times 10^{-6}\,(\Delta m^2/{\rm eV})^2$ {\color{black}(not to be
confused with $L_\star$)}.  
At this stage, collisions become ineffective, and the process enters a mixing-dominated regime, where antineutrinos get resonantly converted
into dark neutrinos.  If the process is fully adiabatic, a maximum final asymmetry $L_{\rm f} = 0.375$ is ultimately generated, corresponding to the
case when all antineutrinos $\nu_1$ are converted into dark neutrinos. 
The exponential stage would then provide the initial value of the asymmetry, corresponding to $L_{\rm i}$ in subsection A, 
needed for the conversion of active to dark neutrinos. 
Notice that, since this asymmetry is generated after neutrino decoupling, 
such asymmetry is not constrained by CMB, since it does not result into any extra radiation compared to the standard case, unless
the value of $\Delta m^2$, and consequently of the resonant temperature, is large enough (above $0.1\,{\rm eV}^2$)
to produce some non-standard BBN and CMB effects \cite{DiBari:2000wd}. 
This option might imply a reduced allowed region in the plane of $\Delta m^2$ versus $\sin^2 2\theta_0$
compared to the case of initial pre-existing asymmetry discussed in subsection A. 
The reason is simply that the onset of the initial exponential growth stage
prior to the resonant conversion is triggered by collisions and these become less efficient at
values of $\sin^2 2\theta_0 \lesssim 10^{-10}$ \cite{DiBari:2001jk}. This reduction would then affect just  marginally
the total (most conservative) allowed region that we show in Section~\ref{sec:IV}. 
\begin{figure*}[t!]
\begin{center}
  \includegraphics[scale=0.385]{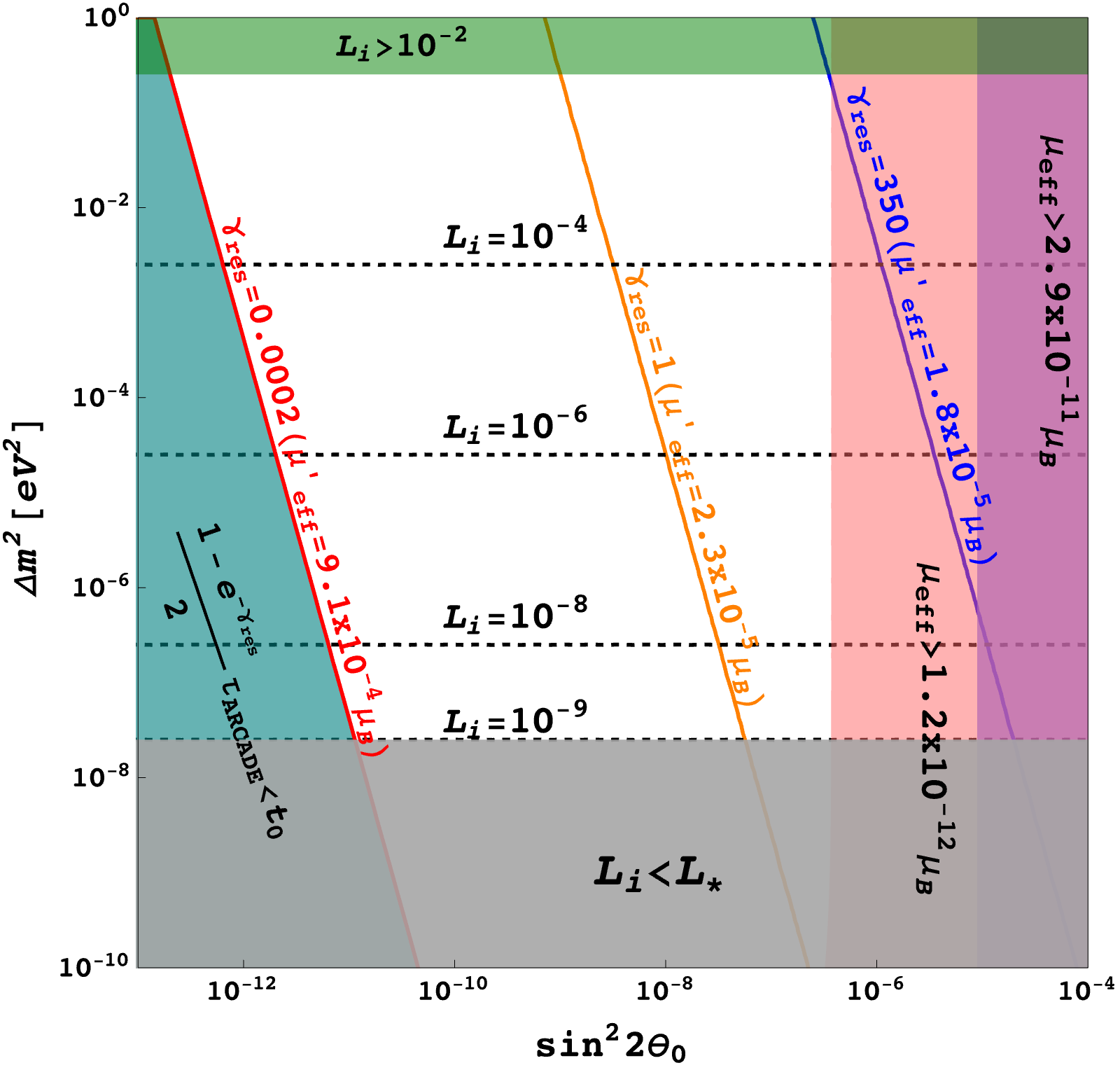}
   \includegraphics[scale=0.4]{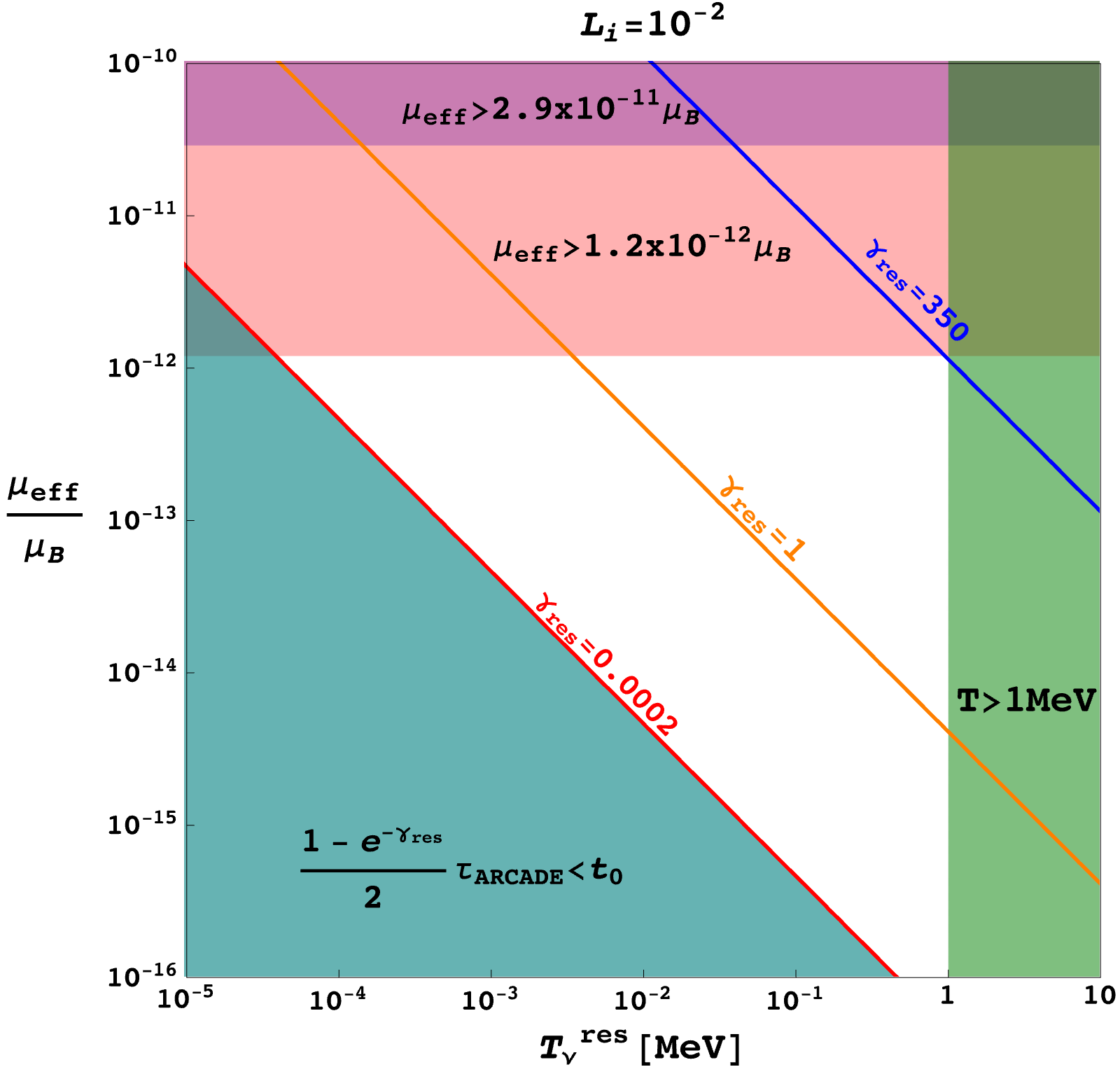}
\end{center}
    \caption{Left panel: Constraints (shaded) and allowed region (white) in the plane of $\Delta m^2$ versus $\sin^2 2\theta_0$. 
    The horizontal dashed lines give an upper bound on $\Delta m^2$ from $T_{\nu}^{\rm res}< 1\,{\rm MeV}$ for the indicated values of $L_{\rm i}$.
    Right panel: Constraints (shaded) and allowed region (white) in the plane of $\mu_{\rm eff}$ versus $T_{\nu}^{\rm res}$.
    We have used conservatively   $\ve=1$.}
    \label{fig:constDm2sin2th0}
\end{figure*}

\section{Dark neutrino decays into dark photons mixed with photons}
\label{sec:III}

In order to explain the ERB, we assume that the dark neutrinos produced by active-to-dark neutrino conversions
in the presence of a pre-existing asymmetry decay into dark fermions $\psi'$ with mass $m_{\psi'}$ and into a superposition of a dark and 
standard photon that we denote by $\gamma'$. The probability that $\gamma'$ is detected as a photon, that can also be regarded as the decay 
branching ration into photons, is denoted by $\varepsilon$. The dark photon is kinematically mixed with the 
standard photon but the kinetic mixing will not play any role, as we will comment. 
The dark fermion is assumed to be quasi-degenerate with $\nu_0$, while we assume the dark photon mass
$m_{\gamma'} \lesssim 10^{-15}\,{\rm eV}$ in order to evade the cosmological constraints from COBE/FIRAS for any value of  
the kinetic mixing parameter $\chi_0$~\cite{Mirizzi:2009iz,Caputo:2020bdy, Garcia:2020qrp, Arsenadze:2024ywr}.\footnote{Here we do not assume the dark photon to be dark matter in which case the bounds on $m_{\gamma'}$ would be more stringent~\cite{Caputo:2021eaa}. Moreover, we do not consider the Juno constraint on $m_{\gamma'}$ that  extends down to $m_{\gamma'}\sim 10^{-18}$ eV~\cite{Yan:2023kdg}, as we lack a detailed understanding of the Jovian magnetosphere and the associated uncertainties.}
We also assume that the decays can be described as fully non-relativistically so that, at the decay, $\gamma'$ has an energy 
$E_{\gamma'}\simeq \D m' \equiv m_0 - m_{\psi'}$. In this way the spectrum of non-thermal photons produced by the
decays of the dark neutrinos and detected at the present time would be described by an effective temperature given exactly by
the expression in Eq.~(\ref{Tgammanth}) multiplied by $\ve$ and also divided by a factor $2 / (1-e^{-{\pi \over 2}\gamma_{\rm res}})$,
taking into account that only relic antineutrinos are, adiabatically or non-adiabatically, converted into dark neutrinos.

The dark fermion radiative decay rate into $\gamma'$ can be related to a dark neutrino effective magnetic moment $\mu_{\rm eff}'$
by an  expression analogous to Eq.~(\ref{Gammanu}):  
\bea\label{Gammanu1}\nonumber
\Gamma_{\nu_{0}  \rightarrow \psi' +\gamma'} & = & {(m^{2}_0 - m^{2}_{\psi'})^3 \over 8\pi \, m_0^3}\, \mu'^{2}_{{\rm eff}}
\simeq {\D m'^{3} \over \pi} \, \mu'^{2}_{{\rm eff}} \\
& \simeq & 42.5\,{\rm s}^{-1}\, \left( {\D m' \over {\rm eV}}\right)^3 \,  \left({\mu'_{{\rm eff}}\over \mu_{\rm B}}\right)^2 \, .
\eea
The important difference is that now $\mu'_{\rm eff}$ does not have direct experimental constraints.
However, because of the active-dark neutrino mixing, the active neutrino still has an effective neutrino magnetic moment
$\mu_{\rm eff} = \sqrt{\ve}\sin^2\theta_0 \mu'_{\rm eff}$ and for this reason the experimental constraints on $\mu_{\rm eff}$ 
still play a role.

\section{Constraints and allowed region}
\label{sec:IV}

Let us now combine all constraints and determine the allowed region in the parameter space of interest. First, we   determine a convenient minimum set of parameters to display all the constraints. We start by imposing that the
non-thermal photons produced by the dark neutrino decays, and mixed with the dark photons, can reproduce the 
ARCADE 2 data. We have then to impose that the dark neutrino lifetime is given by the lifetime determined in~\cite{Dev:2023wel} 
in the case of direct relic neutrino decays (see best fit in Fig.~\ref{fig:bestfit}), that we denote by  $\tau_{\rm ARCADE}(\nu_1 \rightarrow \nu_0+\gamma)$,
shortened by a factor $\ve\,(1- e^{-\gamma_{\rm res}})/2$ to compensate the reduced photon production, explicitly:
\be\label{tauconstraint}
\tau_{\nu_{0}  \rightarrow \psi' +\gamma'} =  {\ve\,(1- e^{-\gamma_{\rm res}})\over 2} \tau_{\rm ARCADE}(\nu_1 \rightarrow \nu_{0}+\gamma) \gtrsim t_0 \,  .
\ee
On the other hand, the lifetime cannot be shorter than $t_0$, as also indicated in (\ref{tauconstraint}), since otherwise the exponential in the decay-law kicks in and suppresses the  photon effective temperature below the ARCADE 2 measured values.  Since the lifetime is the inverse of the decay rate given in
Eq.~(\ref{Gammanu}), the constraint (\ref{tauconstraint}) allows to express $\mu'_{\rm eff}$ in terms of $\gamma_{\rm res}$. Therefore,
for a fixed value of $\gamma_{\rm res}$, one has a fixed value of $\mu'_{\rm eff}$. At the same time $\gamma_{\rm res}$ is expressed in terms
of $\sin^2 2\theta_0$ and $\Delta m^2$ from Eq.~(\ref{gammares}).  

We also have to impose the constraint $T^{\rm res}_{\nu}(L_{\rm i}) \lesssim 1\,{\rm MeV}$, corresponding to a neutrino collisionless regime.\footnote{Here we consider the less constrained option assuming the presence of an initial pre-existing asymmetry $L_{\rm i}$, as discussed in subsection~\ref{sec:IIA}.} 
From Eq.~(\ref{Tnures}) one can see that, 
for a fixed value of $L_{\rm i}$, this constraint implies 
an upper bound on $\Delta m^2$. Higher values of $L_{\rm i}$ correspond to higher allowed values of $\Delta m^2$.
However, one has to impose an upper bound $L_{\rm i} \lesssim 10^{-2}$ from cosmological observations, 
since this would affect both BBN and CMB anisotropies~\cite{Dolgov:2002ab}.  
It is then possible to express all other parameters in terms of 
$\sin^2 2\theta_0$, $\Delta m^2$ and $L_{\rm i}$.  Therefore, we can conveniently show all constraints in the 
plane $\Delta m^2$ versus $\sin^2 2\theta_0$, as shown in the left panel of Fig.~\ref{fig:constDm2sin2th0}
for the most conservative choice $\ve=1$, corresponding to $\gamma'=\gamma$, and highest value of 
$\tau_{\rm ARCADE}(\nu_1 \rightarrow \nu_{0}+\gamma)$ allowed at $99\%$ C.L. (from Eq.~(\ref{ARCADE})).

All constraints (shaded regions) are explicitly indicated. The horizontal dashed lines 
give the upper bound on $\Delta m^2$ for different values of $L_{\rm i}$.  One can also notice the experimental
constraints on $\mu_{\rm eff}$ that basically translate into an upper bound on $\sin^2 2\theta_0$. The orange line 
is the iso-contour line for $\gamma_{\rm res}=1$, marking the border between the adiabatic and the non-adiabatic regime. 
In the right panel of Fig.~\ref{fig:constDm2sin2th0} we also show constraints and allowed region in the plane 
$\mu_{\rm eff}$ versus $T_{\nu}^{\rm res}$ for the maximum allowed value of $L_{\rm i}=10^{-2}$. As one can see, we find a very interesting lower bound
$\mu_{\rm eff}/\mu_{\rm B}\gtrsim 10^{-16}$ which might be testable \cite{Giunti:2024gec}. Taking smaller values of $L_{\rm i}$ moves up this lower bound on $\mu_{\rm eff}$, making it even easier to be tested.

\section{Neutrino oscillation constraints}
\label{sec:V}
 
Having obtained an allowed region in the $\Delta m^2$ versus $\sin^2{2\theta}_0$ plane, we compare it with the existing constraints from neutrino oscillation experiments. Active research is underway on sterile neutrinos with masses around the eV scale, motivated by the excess observed by LSND~\cite{LSND:2001aii},  and the intriguing results reported by  MiniBooNE~\cite{MiniBooNE:2020pnu}, BEST~\cite{Barinov:2021asz}, and IceCube~\cite{IceCubeCollaboration:2024nle}.  However, many other experiments have not found such evidence, leading to significant tension among data sets when combined~\cite{Dentler:2018sju,Hardin:2022muu}.

\begin{figure}[t!]
  \includegraphics[width=0.49\textwidth]{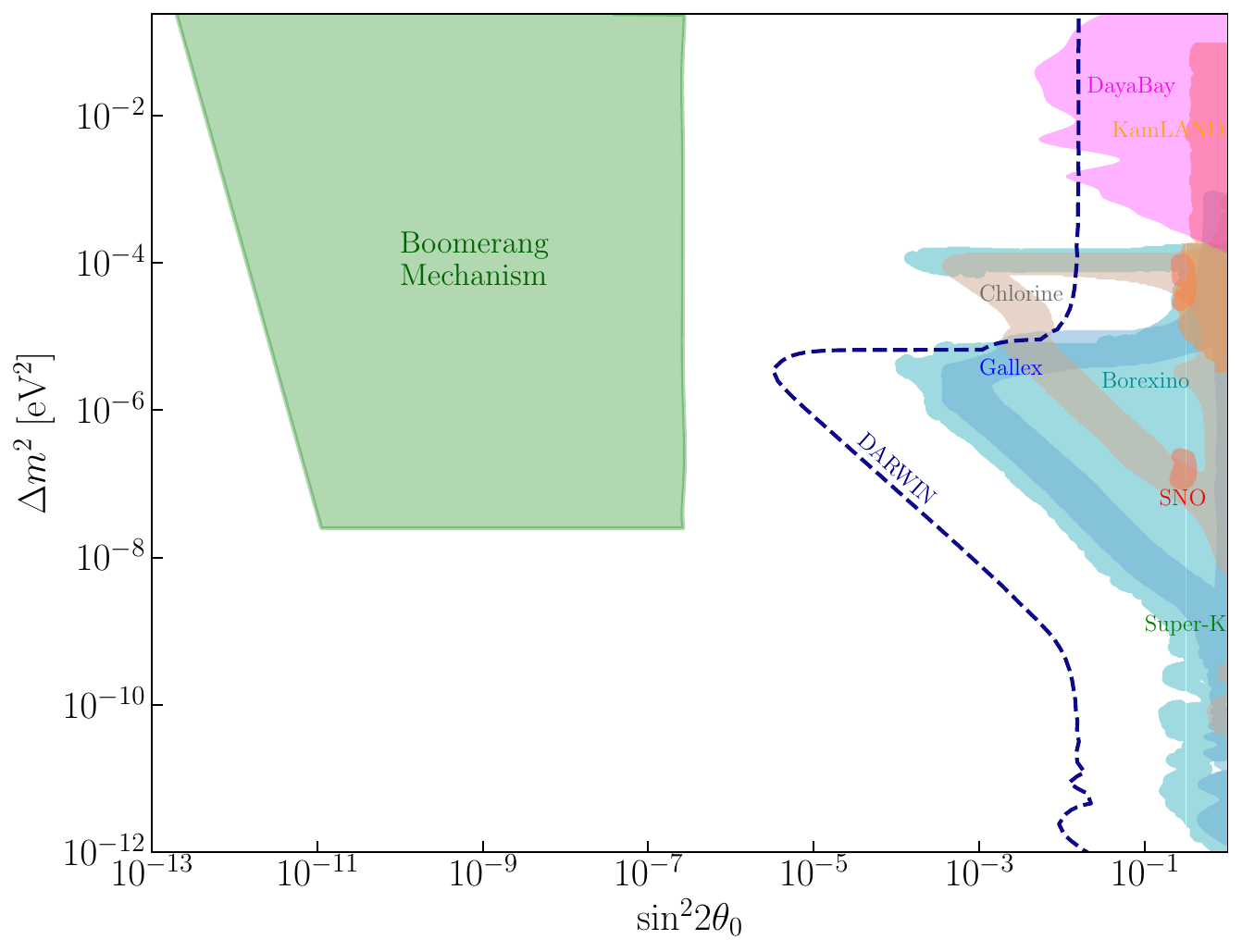}
    \caption{ Region allowed by the boomerang mechanism at 99\% C.L., along with the region excluded by oscillation experiments. We present some of the most stringent bounds on the mixing, including measurements from solar neutrino experiments such as Borexino~\cite{Chen:2022zts}, Gallex~\cite{GALLEX:1998kcz}, Chlorine~\cite{Cleveland:1998nv}, and SNO~\cite{SNO:2002tuh}, Super-Kamiokande~\cite{Super-Kamiokande:2001ljr} as well as reactor experiments like KamLAND~\cite{Chen:2022zts} and Daya Bay~\cite{DayaBay:2024nip}. We also show the future sensitivity from DARWIN~\cite{DARWIN:2020bnc}. }
    \label{fig:osc}
\end{figure}

Beyond the eV-scale sterile neutrino, searches have explored a wide range of mass for oscillations between active and sterile states~\cite{Wolfenstein:1981kw, Petcov:1982ya, Bilenky:1983wt, Kobayashi:2000md, Anamiati:2017rxw, deGouvea:2009fp,Vissani:2015pss, Shoemaker:2015qul, Brdar:2018tce, DeGouvea:2020ang, Rink:2022nvw, Carloni:2022cqz}, but no evidence has been found. For nearly degenerate active and sterile states, solar~\cite{deGouvea:2009fp, Ansarifard:2022kvy, Chen:2022zts}, and reactor data~\cite{DayaBay:2024nip}, constrain $\Delta m^2 \lesssim 10^{-12}\text{eV}^2$ and $\sin^22\theta\lesssim 10^{-4}$, as shown in Fig.~\ref{fig:osc}. The disappearance bounds of muons come from long-baseline experiments~\cite{MINOS:2017cae,NOvA:2024imi} and atmospheric experiments~\cite{Beacom:2003eu}, $\Delta m^2 \lesssim 10^{-4}~\text{eV}^2$ and $\sin^22\theta_0\lesssim 10^{-2}$. Astrophysical neutrinos, with large $L/E$, probe down to $\Delta m^2 \sim 10^{-20}\text{eV}^2$ for maximal mixing~\cite{DeGouvea:2020ang,Martinez-Soler:2021unz,deGouvea:2021ymm,Fong:2024mqz, Dev:2024yrg, Carloni:2025dhv}. 

Figure~\ref{fig:osc} compares the current bounds and future sensitivities with the region predicted by the boomerang mechanism (cf. the left panel of Fig.~\ref{fig:constDm2sin2th0}). Although this work focuses on muon–sterile oscillations, the mechanism applies to all flavors, so we adopt the strongest available limits on the mixing.

Upcoming dark matter experiments such as DARWIN~\cite{DARWIN:2020bnc,deGouvea:2021ymm} are expected to measure the solar neutrino flux from $pp$ fusion via electron scattering. The presence of a light sterile neutrino would modify the electron–neutrino spectrum arriving at Earth. For very small mass splittings ($\Delta m^2 \sim 10^{-12},\text{eV}^2$), sterile neutrinos induce vacuum oscillations along the Earth–Sun baseline~\cite{deGouvea:2021ymm}. As $\Delta m^2$ increases, the oscillation length decreases and matter effects become increasingly relevant, enhancing the sensitivity to smaller mixing angles. For mass splittings around $\Delta m^2 \sim 10^{-6} \text{eV}^2$, mixings as small as $\sin^2 2\theta_{0} \sim 10^{-5}$ could be probed with a 300-ton-year exposure. Reactor experiments, such as JUNO~\cite{Franklin:2023diy}, will also be able to explore the oscillation between active and sterile state for very small mixing angles.

\section{Final remarks}
\label{sec:VI}

\begin{itemize}
\item[(i)] The boomerang mechanism effectively realizes a solution of the ERB mystery
obtained in terms of radiative relic neutrino decays~\cite{Dev:2023wel} but in two stages: in a first early stage
relic antineutrinos (or neutrinos, depending on the sign of $L_{\rm i}$) of one species are converted into 
dark neutrinos (the visible sector throws particles into the dark sector);
in a second stage dark neutrinos decay into dark-standard photon states (dark sector throws back particles into the visible one).\footnote{{\color{black}It should be noticed how the second stage occurs much later (it is basically occurring now). That implies that dark neutrinos are fully non-relativistic when they decay
and their initial momentum distribution, after active-to-dark neutrino conversions, has no impact on the photon spectrum.}}
\item[(ii)] As Fig.~\ref{fig:constDm2sin2th0} shows, it has  a broad variety of phenomenological implications that make it testable in different ways. 
First of all the TMS experiment will soon verify the ARCADE 2 data, the existence of the ERB and the solution
proposed in Ref.~\cite{Dev:2023wel}.  Notice also that this implies some non-standard deviation in the 21 cm cosmological global signal
(see~\cite{Dev:2023wel} for details).
Possible cosmological anomalies in BBN and/or CMB anisotropies might be addressed
by the presence of a large lepton asymmetry~\cite{Burns:2022hkq}. 
Also notice that it implies non-standard relic neutrino background properties
that might be potentially measured~\cite{DiBari:2001jk}. 
Finally, the lower bound we found on $\mu_{\rm eff}$, 
four orders of magnitude below the current upper bound,
will be tested by future experiments~\cite{Giunti:2024gec}. 
\item[(iv)] The monochromatic approximation we used is well justified by numerical solution of density matrix equation 
with a full momentum description~\cite{DiBari:2001jk}. This also grasps 
the variation of the non-adiabaticity parameter with momentum. However, as discussed in Section~\ref{sec:IV}, 
one would have small corrections to the allowed region we have derived. 
\item[(v)] The allowed regions have been determined for the most conservative case $\ve=1$, corresponding to the extreme minimal case
where existence of dark photons would be not necessary. They would clearly shrink for lower values until
they would disappear for a lower bound $\ve \sim 2\times 10^{-4}$. 
\item[(vi)] Having imposed $m_{\gamma'}< 10^{-15}\,{\rm eV}$ suppresses the kinetic mixing since $m^2_\gamma \gg m^2_{\gamma'}$, evading microwave background constraints \cite{Mirizzi:2009iz}. 
Since kinetic mixing is suppressed, we cannot have $\gamma'=\gamma_{\rm dark}$ (i.e., $\ve=0$) and a dynamical generation of a photon component due to a large kinetic mixing parameter $\chi_0$; 
also, since kinetic mixing does not play a role, one could optionally
have $m_{\gamma'} > 10^{-15}\,{\rm eV}$ and negligible kinetic mixing without violating microwave background constraints.
\item[(vii)] Notice that having $\ve$ and $\theta_0 \neq 0$ could induce a non-vanishing neutrino millicharge in addition to an effective 
magnetic moment and this might introduce further constraints to be taken into account \cite{Giunti:2014ixa}. 
However, these constraints are model dependent and we have assumed that the neutrino millicharge is negligible. 
In any case these could be circumvented replacing a constant $\theta_0$ with a temperature dependent effective mixing angle
such that $\theta(T_{\rm res}^\nu)$ coincides with the required values for the mechanism to work shown in the left panel of Fig.~3, 
while for $T\ll T_{\rm res}^{\nu}$  this is sufficiently small to generate a neutrino millicharge in agreement with all constraints. 
Such temperature dependent effective mixing angle can be generated introducing non-standard neutrino interactions yielding 
off diagonal (temperature dependent) terms in the effective Hamiltonian (see Eq.~(\ref{effH})). 
\item[(viii)] We also have to take into account stellar cooling constraints from plasmon decays $\gamma^\star \rightarrow \nu_0 + \psi'$. However,
these can be circumvented coupling the dark fermions to a scalar in a way that either they get a mass higher than plasmon
mass \cite{DeRocco:2020xdt} or that the coupling to photons in stars is suppressed by having a symmetry restoration 
at an energy scale  ${\cal O}({\rm keV})$ such that the scalar vev vanishes \cite{Mohapatra:2006pv}. Interestingly, such a low scale phase transition
has been proposed with independent motivations, within a split-seesaw Majoron model \cite{DiBari:2023upq}.
\item[(ix)] We have not discussed how the boomerang mechanism could be embedded within a full ultraviolet-complete model. For example, a possible direction is offered
by models incorporating quasi-Dirac neutrinos~\cite{Babu:2022ikf, Carloni:2022cqz}. Typically, we have maximal mixing for $\Delta m\ll m$, as $\tan2\theta\sim 2m/\Delta m$, 
but arbitrary mixing is possible for certain textures of the Dirac and Majorana mass matrices, akin to the case of low-scale type-I seesaw~\cite{Kersten:2007vk}. 
Also notice that only within a definite model one could have more specific relation between
$\ve$ and $\chi_0$ and calculate neutrino millicharge and consequent contraints.
\end{itemize}
Before concluding, we would like to highlight that experimentally we will have first important results from the TMS experiment 
\cite{2020SPIE11453E..0TR,Alonso-Arias:2021quq}, that willl test the ERB in the $10$--$20$ GHz range within the next few years. If these should support 
a solution in terms of relic neutrino decays, then the identification of the mechanism responsible for the instability of the relic neutrino
background would become a priority and this would justify dedicated numerical analyses, for example, including a full momentum dependence
and dynamical asymmetry generation. An experimental improvement of the upper bound on the neutrino muon effective magnetic moment 
would then become very important and in case of a positive signal would strongly support the boomerang mechanism. Active-sterile neutrino oscillation
experiments testing very small mixing angles would at this stage become strongly motivated, representing an ultimate test of the mechanism.  
In conclusion, the boomerang mechanism shows that a solution to the ERB in terms of relic radiative neutrino decays is possible.
If the excess is confirmed, this might then provide a direct open window for the exploration of  a new dark sector.

\acknowledgments
We thank Jens Chluba, Andr\'{e} de Gouv\^{e}a, Chee Sheng Fong, Sudip Jana, Jos\`{e} Alberto Rubi\~{n}o Martin, Georg Raffelt, 
Pasquale Serpico, Alexander Studenikin and Anil Thapa for useful discussions. The work of BD was partly
supported by the U.S. Department of Energy under grant No. DE-SC0017987. 
PDB and RR acknowledge financial support from the STFC Consolidated Grant ST/T000775/1. IMS is supported by STFC grant ST/T001011/1.


\bibliography{ref}

\end{document}